\documentclass[twoside,twocolumn,9pt]{article}
\usepackage{extsizes}

\usepackage[super,sort&compress,comma]{natbib} 

\usepackage[version=3]{mhchem}
\usepackage[left=1.5cm, right=1.5cm, top=1.785cm, bottom=2.0cm]{geometry}
\usepackage{balance}
\usepackage{widetext}
\usepackage{times,mathptmx}
\usepackage{sectsty}
\usepackage{graphicx} 
\usepackage{lastpage}
\usepackage[format=plain,ju stification=raggedright,singlelinecheck=false,font={stretch=1.125,small,sf},labelfont=bf,labelsep=space]{caption}
\usepackage{float}
\usepackage{fancyhdr}
\usepackage{fnpos}
\usepackage[english]{babel}
\usepackage{array}
\usepackage{droidsans}
\usepackage{charter}
\usepackage[T1]{fontenc}
\usepackage[usenames,dvipsnames]{xcolor}
\usepackage{setspace}
\usepackage[compact]{titlesec}
\usepackage{bm,amsmath,amssymb,xfrac,float,url,soul}
\usepackage[flushleft]{threeparttable}
\usepackage[colorlinks=true,linkcolor=blue]{hyperref}
\definecolor{cream}{RGB}{222,217,201}
\usepackage{makecell}
\usepackage{commath}
\usepackage[normalem]{ulem}
\usepackage{xcolor}
\usepackage{amssymb}
\floatstyle{plaintop}
\restylefloat{table}

\usepackage{chemformula}
\usepackage{verbatim}
\usepackage{indentfirst}
\usepackage{upgreek}
\usepackage{xcolor}

\colorlet{soulhl}{yellow!30}
\sethlcolor{soulhl}

\usepackage{dblfloatfix}

\usepackage[labelsep=period]{caption}

\graphicspath{{Figures/}}

\begin{document}
\pagestyle{fancy}
\thispagestyle{plain}
\fancypagestyle{plain}{

\renewcommand{\headrulewidth}{0pt}}
\makeFNbottom
\makeatletter
\renewcommand\LARGE{\@setfontsize\LARGE{15pt}{17}}
\renewcommand\Large{\@setfontsize\Large{12pt}{14}}
\renewcommand\large{\@setfontsize\large{10pt}{12}}
\renewcommand\footnotesize{\@setfontsize\footnotesize{7pt}{10}}
\makeatother
\renewcommand{\thefootnote}{\fnsymbol{footnote}}
\renewcommand\footnoterule{\vspace*{1pt} 
\color{cream}\hrule width 3.5in height 0.4pt \color{black}\vspace*{5pt}} 
\setcounter{secnumdepth}{5}
\makeatletter 
\renewcommand\@biblabel[1]{#1}            
\renewcommand\@makefntext[1]
{\noindent\makebox[0pt][r]{\@thefnmark\,}#1}
\makeatother 
\renewcommand{\figurename}{\small{Figure}}
\sectionfont{\sffamily\Large}
\subsectionfont{\normalsize}
\subsubsectionfont{\bf}
\setstretch{1.125}
\setlength{\skip\footins}{0.8cm}
\setlength{\footnotesep}{0.25cm}
\setlength{\jot}{10pt}
\titlespacing*{\section}{0pt}{4pt}{4pt}
\titlespacing*{\subsection}{0pt}{15pt}{1pt}
\fancyfoot{}
%\fancyfoot[LO,RE]{\vspace{-7.1pt}\includegraphics[height=9pt]{head_foot/LF}}
%\fancyfoot[CO]{\vspace{-7.1pt}\hspace{13.2cm}\includegraphics{head_foot/RF}}
%\fancyfoot[CE]{\vspace{-7.2pt}\hspace{-14.2cm}\includegraphics{head_foot/RF}}
\fancyfoot[RO]{\footnotesize{\sffamily{1--\pageref{LastPage} ~\textbar  \hspace{2pt}\thepage}}}
%\fancyfoot[LE]{\footnotesize{\sffamily{\thepage~\textbar\hspace{3.45cm} 1--\pageref{LastPage}}}}
\fancyfoot[LE]{\footnotesize{\sffamily{\thepage~\textbar\hspace{2pt} 1--\pageref{LastPage}}}}
\fancyhead{}
\renewcommand{\headrulewidth}{0pt} 
\renewcommand{\footrulewidth}{0pt}
\setlength{\arrayrulewidth}{1pt}
\setlength{\columnsep}{6.5mm}

%Commented for biblio
%\setlength\bibsep{1pt}
%\makeatletter 

\newlength{\figrulesep} 
\setlength{\figrulesep}{0.5\textfloatsep} 
\newcommand{\topfigrule}{\vspace*{-1pt} 
\noindent{\color{cream}\rule[-\figrulesep]{\columnwidth}{1.5pt}} }
\newcommand{\botfigrule}{\vspace*{-2pt} 
\noindent{\color{cream}\rule[\figrulesep]{\columnwidth}{1.5pt}} }
\newcommand{\dblfigrule}{\vspace*{1pt}
\noindent{\color{cream}\rule[\figrulesep]{\textwidth}{1.5pt}} }
\makeatother

%----commenting--------------------
%\def\myt#1{\textcolor[rgb]{0.5,0,1}{#1}}
\def\myt#1{\textcolor[rgb]{1,0,0}{#1}}
\def\mytfignotdone#1{\textcolor[rgb]{1,0,0}{#1}}

\def\mytnote#1{\textcolor[rgb]{0,0,0}{#1}}

\def\mytfigdone#1{\textcolor[rgb]{0,0,0}{#1}}

\def\gg#1{\textcolor[rgb]{0,0,1}{#1}}
%----------------------------------

%-------- ABSTRACT-----------
\twocolumn[
  \begin{@twocolumnfalse}
\vspace{1cm}
\sffamily
\begin{tabular}{m{0cm} p{16cm} }
%above:\begin{tabular}{m{3.5cm} p{14.5cm} }

%---------TITLE--------------
& \noindent\LARGE{\textbf{Strategies to search for two-dimensional materials with long spin qubit coherence time
%Two-dimensional spin qubit platforms with robust coherence
%High-throughput spin coherence simulations for discovering two-dimensional spin qubit platforms
%Two-dimensional materials with robust spin qubit coherence
%Two-dimensional materials with high spin qubit coherence times
%Identifying two-dimensional materials with high spin qubit coherence
}} \\
%\vspace{0.3cm} & \vspace{0.3cm} \\
%-------------------------------
 & \vspace{1cm} \\
%----------AUTHORS-------------
 & \noindent\large{Michael Y. Toriyama\textit{$^{a \ast}$}, Jiawei Zhan\textit{$^{b}$}, Shun Kanai\textit{$^{c,d,e,f}$}, and Giulia Galli\textit{$^{a,b \ast}$} } \\
 %-------------------------------------
 & \vspace{1cm} \\

 %---------ABSTRACT------------
& \noindent\normalsize{Two-dimensional (2D) materials that can host qubits with long spin coherence time ($T_2$) have the distinct advantage of integrating easily with existing microelectronic and photonic platforms, making them attractive for designing novel quantum devices with enhanced performance.
However, the relative lack of 2D materials as spin qubit hosts, as well as appropriate substrates that can help maintain long $T_2$, necessitates a strategy to search for candidates with robust spin coherence.
Here, we develop a high-throughput computational workflow to predict the nuclear spin bath-driven qubit decoherence and $T_2$ in 2D materials and heterostructures.
We initially screen \mytnote{1173} 2D materials and find \mytnote{190} monolayers with $T_2 > 1$ ms, higher than that of naturally-abundant diamond.
We then construct \mytnote{1554} lattice-commensurate heterostructures between high-$T_2$ 2D materials and select 3D substrates, and we find that $T_2$ is generally lower in a heterostructure than in the bare 2D host material; however, low-noise substrates (such as \ch{CeO2} and CaO) can help maintain high $T_2$.
To further accelerate the material screening effort, we derive analytical models that enable rapid predictions of $T_2$ for 2D materials and heterotructures.
The models offer a simple, yet quantitative, way to determine the relative contributions to decoherence from the nuclear spin baths of the 2D host and substrate in a heterostructural system.
By developing a high-throughput workflow and analytical models, we expand the genome of 2D materials and their spin coherence times for the development of spin qubit platforms.
}\\ 
 &\vspace{0.3cm} \\
\end{tabular}
\end{@twocolumnfalse} \vspace{0.6cm}
]

%----------	FOOTNOTE ------------
\renewcommand*\rmdefault{bch}\normalfont\upshape
\rmfamily
\section*{}
\vspace{-1cm}
\footnotetext{
$^{a}$Materials Science Division, Argonne National Laboratory, Lemont, IL 60439, USA. \\ 
$^{b}$Pritzker School of Molecular Engineering, University of Chicago, Chicago, IL 60637, USA. \\
$^{c}$Laboratory for Nanoelectronics and Spintronics, Research Institute of Electrical Communication, Tohoku University, Sendai 980-8577, Japan. \\
$^{d}$WPI-Advanced Institute for Materials Research (WPI-AIMR), Tohoku University, Sendai 980-8577, Japan. \\
$^{e}$National Institutes for Quantum Science and Technology, Takasaki 370-1207, Japan. \\
$^{f}$Division for the Establishment of Frontier Sciences, Tohoku University, Sendai 980-8577, Japan. \\
\textit{$^*$E-mail: mtoriyama@anl.gov, gagalli@uchicago.edu}
}

% -----------------------------
% Journals
%   - npj 2D Materials and Applications (special collection on 2D materials for QIS: https://www.nature.com/collections/fiacfeiicf)
%   - PRX (special issue on 2D materials: https://journals.aps.org/prx/collections/prx-2dmaterials)
%   - Appl Phys Lett
%   - Phys Rev Mater
%   - Phys Rev B
%   - Phys Rev Lett
%   - PRX Quantum
% -----------------------------

%------------------------------
\section{Introduction}

Spin defects in solid-state materials offer a promising platform for realizing qubit-based technologies. However, the loss of phase information between the defect's levels due to the environment can significantly reduce the reliability of a qubit and, as a result, limit the performance of a quantum device. In the regime where the dephasing rate significantly exceeds the relaxation rate, the qubit's decoherence is characterized by the the time it takes to lose memory of its phase (called decoherence time $T_2$).
%Depending on the pulse sequences used experimentally to interrogate the qubit, including the Ramsey and Hahn echo sequences, one may measure varied coherence times $T_2$. 
Decoherence can occur due to a number of reasons. In solids, magnetic noise from nuclear isotopes can play a major role. As a result, the choice of the host material, which determines the nuclear spin environment of the qubit, is an important design consideration that strongly affects the performance of quantum devices.

Many three-dimensional (3D) materials have been studied as spin qubit hosts, including diamond, silicon carbide, and oxides,  which are systems with relatively low magnetic noise. However, in many applications, in particular quantum sensing, the spin defect needs to be placed near the surface of the material, where dangling bonds and surface defects can introduce additional unwanted noise leading to decoherence.\cite{Janitz2022_Review} Two-dimensional (2D) materials, on the other hand, offer distinct advantages over 3D qubit hosts, for example the density of nuclear isotopes is inherently lower than in 3D materials, thus limiting the decoherence from magnetic noise.\cite{Ye2019_SpinCoherence_2D, Onizhuk2021_Substrate} In addition, if 2D monolayers can be exfoliated from 3D van der Waals-bonded layered systems, the decoherence due to dangling bonds can be greatly limited. Atomically-precise creation of spin defects on monolayers has been demonstrated using, e.g., scanning tunneling microscopy,\cite{Schuler2020_WS2, Cochrane2021_WS2_STM, Thomas2024_WS2_QuantumDefect} enabling precise control over functional properties. Further, 2D materials can be integrated with microelectronic and photonic platforms,\cite{Fang2024_Review_vdW_Sensing} possibly enabling the design of devices with enhanced capabilities and/or novel functionalities.

Studies of qubits in 2D materials have so far been limited to few platforms, such as h-BN, transition metal dichalcogenides, and recently \ch{GeS2}.\cite{Zhang2020_Review_DefectQubits, Montblanch2023_Review_Layered, 2024_Liu_GeS2, 2025_Vaidya_GeS2} In order to realize quantum devices equipped with specific functions, a substantially wider selection of 2D host materials is desirable. In a previous report, the Hahn echo $T_2$ times of 69 spin defects in 45 2D host materials were calculated,{\cite{Sajid2022_2D_Coherence}} offering important insights into the spin coherence properties of 2D materials. However, to gain a deeper understanding of these properties, and include the effect of the substrate over which 2D materials reside, a much broader set of systems need to be considered.  Critically, the existing literature lacks data-driven insights into substrate-induced decoherence effects in heterostructural qubit hosts. In addition, while a model of $T_2$ was recently developed to rapidly screen thousands of potential 3D qubit host materials,{\cite{Kanai2022_ScalingLaw}} no such model is available for 2D systems. Such a model can accelerate predictions of $T_2$ for large sets of materials, and even possibly enable AI-assisted design of materials with robust spin qubit coherence.

Here, we develop and implement a computational strategy to search for 2D materials and heterostructural systems with appropriate substrates which can maintain long coherence times. We compute the spin-echo $T_2$ time using the cluster correlation expansion (CCE) approach.\cite{Onizhuk2025_Review} A key capability developed in our work is an automated framework to perform CCE simulations with little manual intervention, enabling large-scale predictions of $T_2$ for a vast set of materials. We initially apply this approach to \mytnote{1173} 2D materials sourced from the MC2D database, finding \mytnote{190} monolayers with $T_2 > 1$ ms. We find that 2D materials with the 15 highest predicted $T_2$ all have a band gap greater than 2 eV, suggesting that they can host defect-induced electronic levels within their band gap to realize a spin qubit. We then construct \mytnote{1554} heterostructures between high-$T_2$ monolayers and 3D substrates with commensurate lattices. Our data-driven approach correctly predicts that qubit coherence can be limited by either the 2D host material or the substrate, depending on the material with the noisier nuclear spin environment. Using our data, we propose physically-motivated models of $T_2$ for 2D materials and heterostructures, enabling rapid predictions of coherence times from only structural information. We apply the model to 2D materials sourced from several databases, allowing us to expand our search space by 4740 additional monolayers. 
%All of our data, including structures and calculated $T_2$, are provided as Supplemental Data in Ref. \myt{insert}.

Prior to describing our results, we make a few important remarks on nomenclature. First, we note that monolayers considered in this work are not all strictly two-dimensional and are instead few atoms thick. However, we will refer to them as ``2D materials'' and ``monolayers'' interchangeably to reflect their small thickness, which is a key property than enables long coherence times due to the low nuclear spin density.\cite{Ye2019_SpinCoherence_2D} Second, when it is stated that a material ``exhibits high/long $T_2$'', this refers to the fact that the material can \textit{accommodate a spin qubit with long $T_2$}. While $T_2$ is, strictly speaking, a property of the spin defect/qubit, we attribute it to the host material in this study to reflect the decoherence effects through hyperfine interactions with the nuclear spin bath.

\section{Results}

\begin{figure}[!t]
    \centering
    \includegraphics[width=0.48\textwidth]{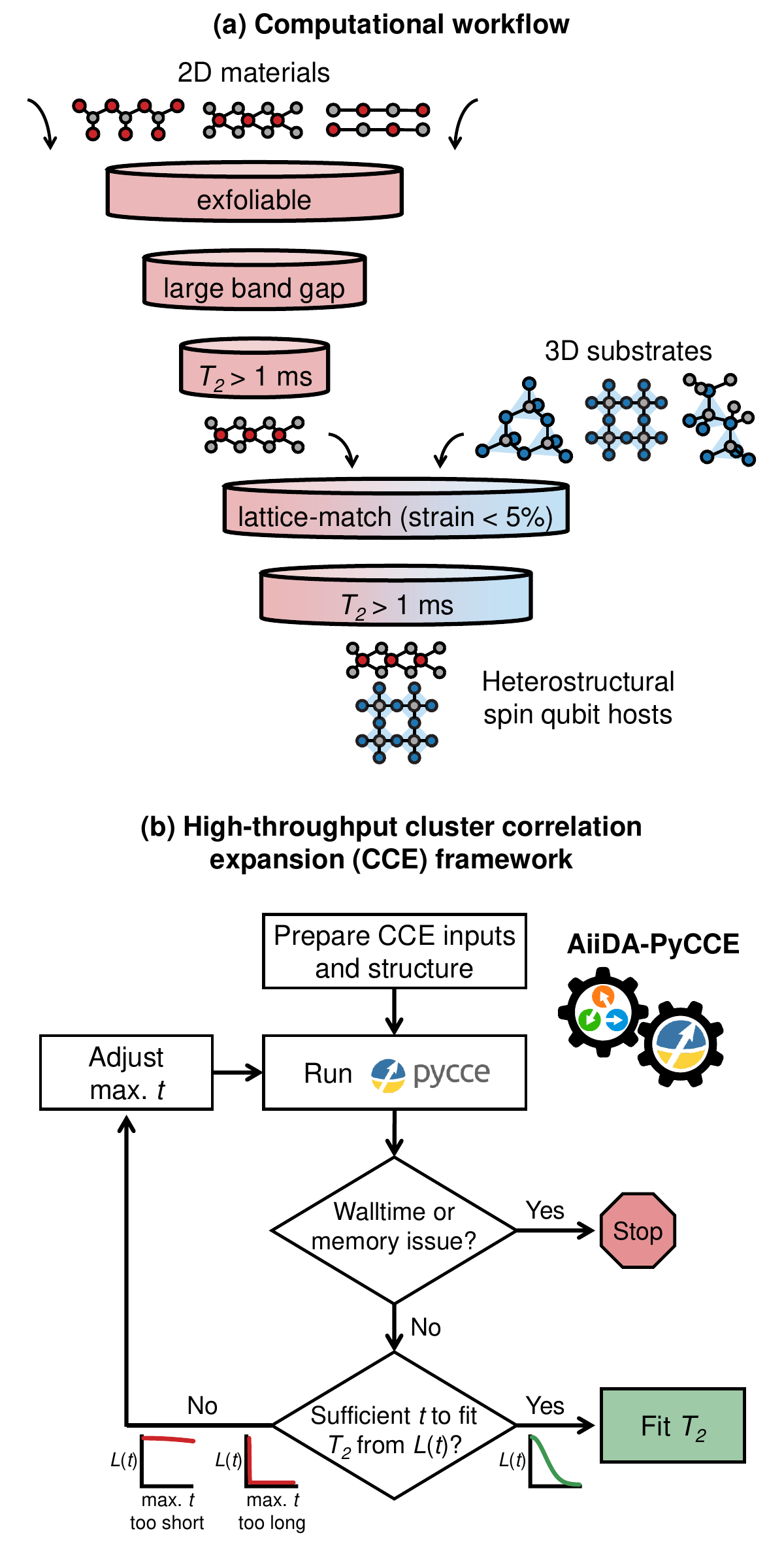}
    \caption{
    \textbf{Computational methodology.}
    (a) The workflow for screening 2D materials and heterostructures with long coherence times ($T_2$) adopted in this work. Promising 2D materials with high $T_2$ times are identified initially, followed by the creation and screening of heterostructures.
    (b) The automated methodology for performing cluster correlation expansion (CCE) simulations to calculate $T_2$ devised in this work.
    }
    \label{Fig:Workflow}
\end{figure}

\subsection{High-throughput workflow} \label{Sec:HT_Workflow}

We show in Figure \ref{Fig:Workflow}a our tiered computational workflow for identifying 2D host materials and heterostructures with long Hahn-echo coherence time ($T_2$). Broadly, the workflow is composed of two stages: we first identify 2D materials with high $T_2$, which are then combined with 3D substrates and screened for high-$T_2$ heterostructures. This is motivated by the idea that high $T_2$ for the monolayer is essential, because the substrate will introduce additional noise to the qubit and lead to lower $T_2$ in a heterostructural system. We show in Section \ref{Sec:Heterostructure_T2} that this hypothesis is largely valid, where calculations show that $T_2$ of heterostructures are generally lower than that of the base 2D host material.

\textbf{Description of 2D materials set.}
With the rise of material informatics, several 2D materials databases have been developed in recent years.\cite{Mounet2018_MC2D, Campi2023_MC2D, Zhou2019_2DMatPedia, Haastrup2018_C2DB, Gjerding2021_C2DB, Wang2023_SymmBased} For the purposes of experimental realization, it is important to consider realistic 2D materials. While various thermodynamic arguments can be made to demonstrate the stability of a hypothetical 2D material, perhaps a more effective strategy is to consider monolayers that can be exfoliated from \emph{known} materials.

Therefore, in our work, we consider 2D materials that are curated by the MC2D database.\cite{Mounet2018_MC2D, Campi2023_MC2D, MC2D_Data} This database offers the distinct advantage of compiling monolayers that are isolated from 3D layered materials that are \emph{experimentally-known}, and are sourced from experimental crystal structure databases such as ICSD,\cite{ICSD_1, ICSD_2} COD,\cite{COD_1, COD_2} and MPDS.\cite{MPDS} Key properties that indicate whether a monolayer can be exfoliated from its corresponding 3D parent phase, such as the binding energy, are calculated and recorded in the database. Although many of the monolayers are still hypothetical, in that they have not yet been experimentally realized in monolayer form, comparing binding energies to those of monolayers that have been successfully realized (e.g., h-BN and graphene) enables us to identify those that can potentially be exfoliated in experiments. In particular, we consider materials that have DFT binding energy that is less than 130 meV/\AA$^2$ when calculated using the DF2-C09 and rVV10 functionals, which are considered either ``easily'' or ``potentially'' exfoliable in Ref. \citenum{Mounet2018_MC2D}.

\begin{table*}[!t]
\centering
\caption{2D materials with the \mytnote{15} highest spin coherence times ($T_2$) identified using our computational workflow. The monolayers are sourced from the Materials Cloud 2D crystals database (MC2D) with their IDs listed, the coherence time is calculated from cluster correlation expansion (CCE) simulations, and the band gap is predicted using the screened-exchange range-separated hybrid (SE-RSH) functional. For each 2D host material, there is a corresponding 3D layered phase that is experimentally known; we list the database and ID number for each parent phase.
}
\begin{tabular}{ ccccccc }
\Xhline{0.5\arrayrulewidth}
&&& \\
2D host & $T_2$ (ms) & $E_g$ (eV) & MC2D ID & Exp. database & Exp. ID & Notes \\
&&& \\
\Xhline{0.5\arrayrulewidth}
&&& \\
\ch{WS2} & 35.4 & 2.61 & 234 & COD & 9009145 & Well-known\cite{Cochrane2021_WS2_STM, Thomas2024_WS2_QuantumDefect} \\
\ch{Au2Se3O10} & 23.5 & 4.17 & 1214 & COD & 4311594 & \\
\ch{Au2Se2O7} & 16.2 & 4.16 & 1205 & COD & 7209412 & \\
\ch{PdSO4} & 14.8 & 3.81 & 2224 & ICSD & 79559 & Successfully exfoliated\cite{Alkathiri2022_PdSO4}\\
\ch{Hg2GeO4} & 14.2 & 5.6 & 1451 & ICSD & 26340 & \\
\ch{AgCO2} & 14.0 & 5.32 & 377 & ICSD & 109600 & \\
\ch{KNO3} & 11.7 & 8.23 & 1667 & ICSD & 384 & \\
\ch{KAgTeS3} & 11.6 & 3.49 & 2040 & MPDS & S1804089 & \\
\ch{K2Ag2GeS4} & 11.2 & 4.32 & 922 & ICSD & 170843 & \\
\ch{AuSe} & 10.6 & 2.07 & 14 & COD & 1510294 & $\beta$-phase \\
\ch{PbSO3} & 10.5 & 5.75 & 656 & COD & 9009622 & \\
\ch{K2Ge2S5} & 10.1 & 4.89 & 1850 & ICSD & 411027 & \\
\ch{GeS2} & 9.9 & 3.19 & 1951 & MPDS & S1831556 & High-pressure tetragonal phase \\
\ch{PbSeO3} & 9.8 & 6.11 & 1279 & COD & 1526029 & \\
\ch{GeS} & 9.6 & 2.78 & 159 & COD & 9008784 & \\
&&& \\
\end{tabular}
\label{Table:T2_2D}
\end{table*}
%--------------------

Additionally, realizing spin qubits materials with wide band gaps is ideal, because these materials can potentially accommodate localized defect states in the band gap which do not interfere with host states.\cite{Weber2010_QuantumComputing_Defects} The band gap of each monolayer in the MC2D database is predicted using the generalized gradient approximation (GGA), which typically underestimates the measured band gap of a material. To ensure that we consider systems with sufficiently wide band gaps without missing potential spin qubit hosts, we consider materials with a GGA band gap greater than 0.5 eV in our work. We then refine our band gap predictions of monolayers with the highest $T_2$ times by using the screened-exchange range-separated hybrid (SE-RSH) functional, which yield higher accuracy results for solids and interfaces\cite{Zhan2023_SERSH, Zhan2025_SERSH_MetalOxides}

\textbf{Automated framework for performing cluster correlation expansion simulations.}
A key capability gap addressed in this work is a high-throughput framework to calculate the $T_2$ coherence times of a large set of materials. We use the cluster correlation expansion (CCE) approach, as implemented in the open-source PyCCE code,\cite{Onizhuk2021_PyCCE} to simulate the decoherence dynamics of a spin qubit in the presence of nuclear isotopes in a host material. The CCE method allows us to calculate the coherence function $\mathcal{L}(t)$ of a spin qubit, from which the $T_2$ time can be extracted by fitting $\mathcal{L}(t)$ to a stretched exponential form, $\mathcal{L}(t) = \exp\left[-\left(t/T_2\right)^n\right]$. More details can be found in Section \ref{Sec:Methods}.

For each CCE calculation, the maximum time and number of time steps to simulate $\mathcal{L}(t)$ must be decided. Ideally, the number of time steps will be sufficiently large to obtain enough information on $\mathcal{L}(t)$, but an arbitrarily large number of steps would lead to walltime and/or memory issues. Moreover, the maximum simulation time cannot be too short nor arbitrarily long, otherwise the simulated $\mathcal{L}(t)$ will be incomplete or would decay too early within the simulated time domain, respectively (Figure \ref{Fig:Workflow}b). In either case, one cannot obtain a reliable fit of $T_2$. As a result, a workflow manager that can track each simulation is needed, especially when handling a large set of materials.

Here, we built a software based on the AiiDA workflow manager\cite{Huber2020_AiiDA, Uhrin2021_AiiDA} to run PyCCE in a high-throughput manner (Figure \ref{Fig:Workflow}b). The software, which we call AiiDA-PyCCE,\cite{AiiDA_PyCCE_Github} initializes each calculation from only the structure of a material and basic CCE inputs. A key feature of this software is the ability to automatically check each simulation to ensure it ran for an appropriate time (Figure \ref{Fig:Workflow}b), thus enabling reliable fits of $T_2$ with no manual intervention. Using our high-throughput framework, we identify materials with $T_2 > 1$ ms (Figure \ref{Fig:Workflow}). We choose this bound to be similar to that of naturally-abundant diamond, which has a coherence time of $T_2 \sim 0.9$ ms when calculated using similar CCE methods.\cite{Kanai2022_ScalingLaw}

Once 2D materials with long $T_2$ time have been identified, heterostructures are created between them and 3D substrates cleaved along different planes (Figure \ref{Fig:Workflow}a). The list of substrates considered in this study is given in \mytfigdone{Table S1}. We create heterostructures where the interlayer separation between the 2D host and substrate is set to the sum of van der Waals radii{\cite{Alvarez2013_vdW_Radii}} of the atoms at the interface. We also ensure that the lattices are commensurate by applying a strain no greater than 5\% to the 2D material, using an algorithm proposed by Zur and McGill{\cite{Zur1984_LatticeMatching}} and implemented in the MPInterfaces code.{\cite{Mathew2016_MPInterfaces}} The lattice commensurability requirements depend on the type of bonding at the heterointerface. If the 2D material is chemically bonded (e.g. covalently) to the substrate, then the lattice will likely become strained to accommodate the interfacial bonds, thus enforcing lattice commensurability. On the other hand, if the 2D material is weakly bound to the substrate through van der Waals forces, we can expect the lateral strain on the monolayer to be minimal, even when the lattices themselves are incommensurate.{\cite{Utama2013_Review_vdWE}} However, lattice incommensurability leads to technical problems for \textit{ab initio} methods with periodic boundary conditions, where a repeatable unit cell must be defined. Lattice-matching ensures that each heterointerface can be defined within a cell of finite size, thus enabling \textit{ab initio} calculations of these heterostructures. We do not relax the interfacial structure to ensure that the workflow can be executed in a high-throughput manner. We show in Section \ref{Sec:Heterostructure_T2} and the Supplemental Information that this is largely justified by the relative insensitivity of the nuclear spin bath-induced decoherence to specific details of the interfaces considered in our study, such as the interlayer separation, strain, and atomic reconstruction.

\begin{figure*}[!b]
    \centering
    \includegraphics[width=\textwidth]{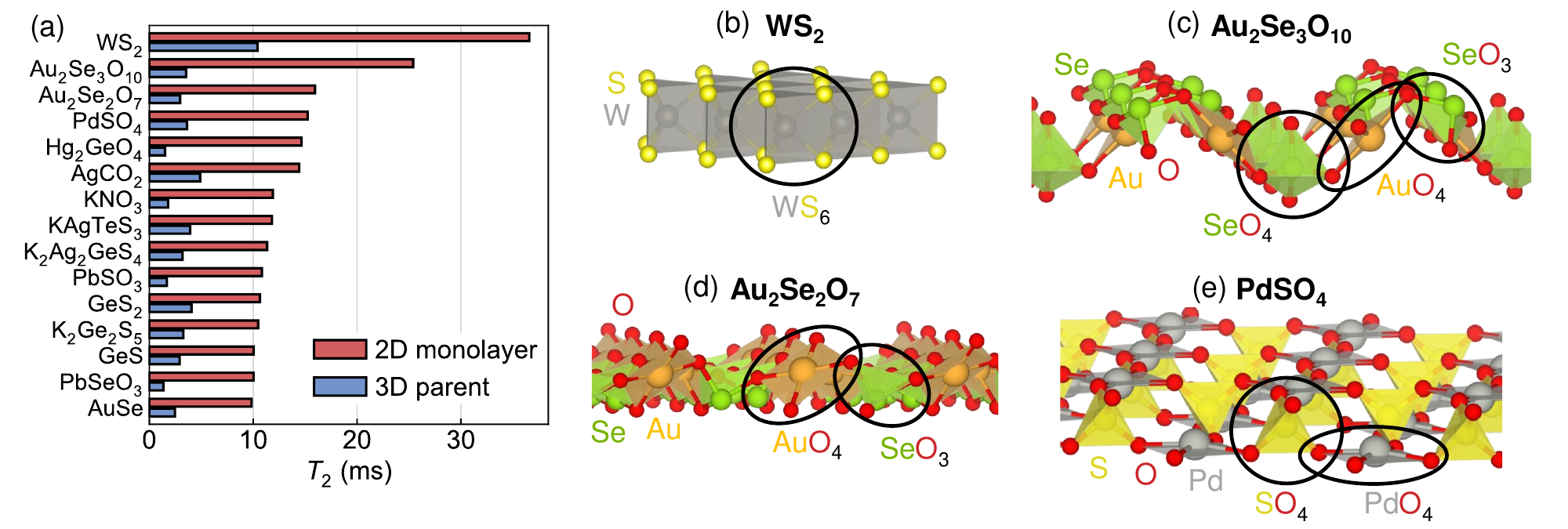}
    \caption{
    \textbf{2D materials with high coherence time.}
    (a) Comparison of spin coherence time ($T_2$) for qubits implanted in 2D monolayers (red) and corresponding 3D parent phases (blue). 
    2D materials can generally accommodate spin qubits with longer coherence time than their 3D parent phase.
    Structures of monolayers with the highest $T_2$ are shown, namely (b) \ch{WS2}, (c) \ch{Au2Se3O10}, (d) \ch{Au2Se2O7}, and (e) \ch{PdSO4}.
    }
    \label{Fig:T2_HT}
\end{figure*}

\subsection{Coherence time $T_2$ of 2D materials}

Out of \mytnote{1173} monolayers, we find \mytnote{190} monolayers that exhibit $T_2 > 1$ ms using our high-throughput workflow. The top \mytnote{15} candidates are shown in Table \ref{Table:T2_2D}, along with their band gaps calculated using the screened-exchange range-separated hybrid (SE-RSH) functional,\cite{Zhan2023_SERSH, Zhan2025_SERSH_MetalOxides} the ID in the MC2D database, and the experimental details of the 3D parent phase from which the monolayer can be exfoliated. Although we only consider monolayers with GGA band gap greater than 0.5 eV in our screening (see Section \ref{Sec:HT_Workflow}), our calculations with the SE-RSH functional show that monolayers with the 15 highest $T_2$ have band gaps greater than 2 eV (Table \ref{Table:T2_2D}). It may therefore be possible to realize a spin qubit with a localized two-level system through doping in each of these monolayers. For reference, \ch{WS2}, which has been used as a spin qubit platform,\cite{Schuler2020_WS2, Cochrane2021_WS2_STM, Thomas2024_WS2_QuantumDefect} has a predicted band gap of 2.61 eV in monolayer form (Table \ref{Table:T2_2D}), in agreement with other quasiparticle calculations (2.9--3.1 eV){\cite{Shi2013_WS2_QP, Chen2022_2D_Fock}} and experimental values (2.4--2.7 eV).{\cite{Chernikov2014_WS2_Gap, Zhu2015_WS2_Gap}}

Our results generally support the conclusion that 2D materials can accommodate longer $T_2$ times than 3D materials. By comparing the $T_2$ times of 2D monolayers against their corresponding 3D parent phases, we find that 2D materials generally exhibit longer $T_2$ times (Figure \ref{Fig:T2_HT}a). This can be attributed to their low dimensionality and, as a result, inherently lower nuclear spin density.

The monolayer with the highest predicted $T_2$ is \ch{WS2} ($T_2 = 35.4$ ms, Table \ref{Table:T2_2D}), which has been tested as a spin qubit platform.\cite{Schuler2020_WS2, Cochrane2021_WS2_STM, Thomas2024_WS2_QuantumDefect} The high $T_2$ can be attributed to the relatively low $g$-factor of $^{183}$W (0.24) and the low nuclear spin density of $^{33}$S (0.76\%). In fact, we find that many high-$T_2$ materials are oxides or sulfides, with some even being heteroanionic with two or more chalcogen species (Table \ref{Table:T2_2D} and \mytfigdone{Figure S1}), owing to the low natural abundance of isotopes with nonzero spin, specifically $^{17}$O (0.038\%) and $^{33}$S (0.76\%). A similar conclusion has been drawn for 3D solids.\cite{Kanai2022_ScalingLaw}

The Au-oxyselenide monolayers \ch{Au2Se3O10} ($T_2 = 23.5$ ms) and \ch{Au2Se2O7} ($T_2 = 16.2$ ms) exhibit the second- and third-highest $T_2$ times in our list. Although $^{197}$Au is a spinful nuclear species ($S = 3/2$) with 100\% natural abundance, the relatively low nuclear $g$-factor (0.097) and large interatomic separation enables high $T_2$ in these host materials. The oxygen-rich chemical makeup is also beneficial for spin coherence due to the low natural abundance of the $^{17}$O isotope. Note that there are corner-sharing units of square-planar AuO$_4$ and units of SeO$_3$ in both structures (Figures \ref{Fig:T2_HT}c and \ref{Fig:T2_HT}d), as well as near-tetrahedral units of SeO$_4$ in \ch{Au2Se3O10} (Figure \ref{Fig:T2_HT}c). As tabulated in the MC2D database, the binding energies of the two compounds ($E_b = 27.6$ meV/\AA$^2$ for \ch{Au2Se3O10} and $E_b = 27.3$ meV/\AA$^2$ for \ch{Au2Se2O7}) are comparable to that of \ch{PtSe2} ($E_b = 29.6$ meV/\AA$^2$), which has been successfully exfoliated through mechanical\cite{Avsar2020_PtSe2_MechanicalExf, Aslam2024_PtSe2_MechanicalExf} and electrochemical\cite{Cho2023_PtSe2_Electrochemical} means. This suggests that both Au-oxyselenide monolayers can also be exfoliated from their respective 3D layered parent phases.

We find that \ch{PdSO4} is another candidate spin qubit host with high $T_2$ (14.8 ms), which can be attributed to the low nuclear spin densities of $^{17}$O and $^{33}$S, as well as the relatively low nuclear $g$-factor of $^{105}$Pd (-0.257), an isotope that exists with 22.33\% natural abundance. The structure is embedded with units of tetrahedral SO$_4$ and square-planar PdO$_4$ (Figure \ref{Fig:T2_HT}e).\cite{Dahmen1994_PdSO4} Sheets of \ch{PdSO4} with $\sim3$ nm thickness were recently synthesized through liquid-phase exfoliation,\cite{Alkathiri2022_PdSO4} demonstrating potential for near-term testing as a spin qubit host. This is perhaps not surprising given the low binding energy of the material ($E_b = 8.54$ meV/\AA$^2$), which is lower than that of graphene ($E_b = 20.29$ meV/\AA$^2$) and \ch{MoS2} ($E_b = 21.56$ meV/\AA$^2$).\cite{Mounet2018_MC2D, Campi2023_MC2D}

$\beta$-\ch{GeS2} (space group $P2_1/c$) has recently been gaining traction as a potential spin qubit host at ambient conditions.\cite{2024_Liu_GeS2, 2025_Vaidya_GeS2} Although monolayer $\beta$-\ch{GeS2} is not listed in the MC2D database, we find that a monolayer of the high-pressure tetragonal phase of \ch{GeS2} (space group $P4_2/nmc$) has $T_2 = 9.9$ ms (Table \ref{Table:T2_2D}). Separately, we isolate a monolayer from the 3D layered phase of $\beta$-\ch{GeS2} found in the Materials Project database (mp-572892) and find that it possesses $T_2 = 12.5$ ms, which is slightly higher than the $T_2$ of the tetragonal phase. Our predicted $T_2$ is orders of magnitude larger than the experimentally-measured $T_2$ (< 20 $\mu$s),\cite{2024_Liu_GeS2, 2025_Vaidya_GeS2} indicating that the qubit coherence is limited by a noise source other than the nuclear spin bath, such as dangling bonds that generate electric noise.

Our study elucidates the vast chemical diversity of 2D materials that can host spin qubits with long $T_2$.
For example, 4 out of the 15 materials with the highest $T_2$ contain potassium in its composition (Table \ref{Table:T2_2D}); despite being an element with 100\% spinful isotopes, the high $T_2$ is attributable to the relatively low $g$-factor (e.g. 0.26 for $^{39}$K) and dilute presence in each composition.
Nonetheless, it is interesting to note some common structural motifs in high-$T_2$ monolayers, such as square-planar units of transition metal oxides (Figures {\ref{Fig:T2_HT}}c-{\ref{Fig:T2_HT}}e). A qubit's electronic structure and, therefore, properties, are influenced by the coordination symmetry of the defect center. For example, defect centers that are inversion-symmetric, such as the SiV$^-$ center in diamond,{\cite{Sipahigil2014_SiV}} do not possess a dipole moment and are therefore less sensitive to electric field noise compared to centers that lack inversion symmetry. As a result, the choice of dopant for the spin defect center, as well as the choice of host material, can enable the design of quantum devices with a wide range of capabilities. As we discuss in detail in Section \ref{Sec:Discussion}, the chemical richness and diversity of structural motifs embedded in these host structures can potentially enable a variety of spin defect centers with specific electronic structures and, therefore, desirable properties.

\begin{figure}[!t]
    \centering
    \includegraphics[width=0.5\textwidth]{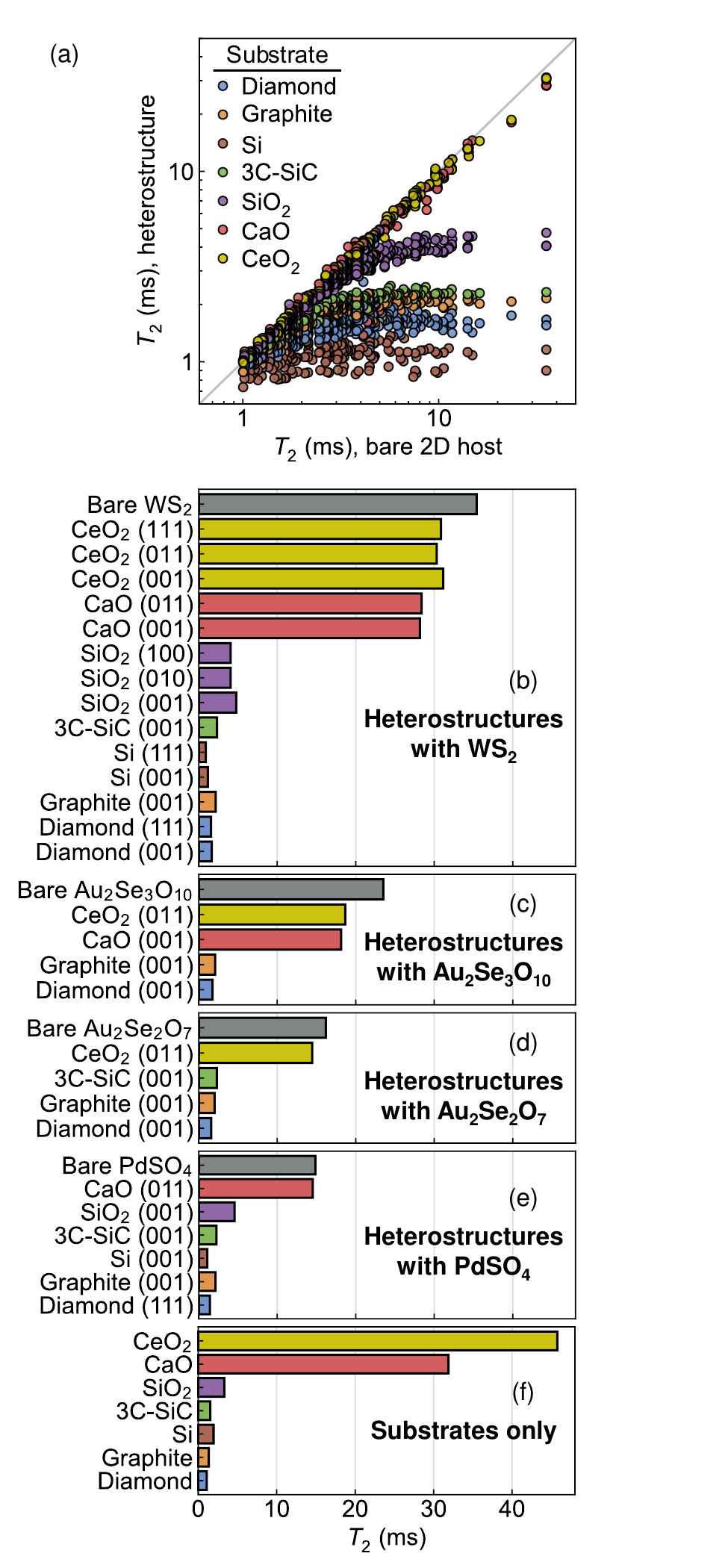}
    \caption{
    \textbf{Coherence time of heterostructures.}
    (a) Spin coherence time ($T_2$) of heterostructures compared to $T_2$ of the bare 2D host material. The color indicates the substrate material.
    (b-e) $T_2$ of heterostructures with \ch{WS2}, \ch{Au2Se3O10}, \ch{Au2Se2O7}, and \ch{PdSO4}. The colored bars represent heterostructures with the substrate indicated along the vertical axis, and the gray bar represents the bare 2D host.
    (f) $T_2$ of the bare substrate materials, without the 2D material.
    }
    \label{Fig:T2_Heterostructures}
\end{figure}

\subsection{Coherence time $T_2$ of heterostructures} \label{Sec:Heterostructure_T2}

Next, we consider heterostructures between 2D materials with $T_2$ > 1 ms and select 3D substrates cleaved along crystallographic planes. We create lattice-matched heterostructures by straining the 2D material by no more than 5\%, which generates \mytnote{1554} heterostructures. We use the high-throughput procedure implemented in the AiiDA-PyCCE code to compute $T_2$ for all heterostructures, where the qubit is placed in the 2D host material. It is important to note that we only account for decoherence driven by the nuclear spin baths of the substrate and 2D host material. While other sources of decoherence may exist, such as surface roughness and dangling bonds at the host--substrate interface,{\cite{Janitz2022_Review}} the nuclear spin bath-limited $T_2$ can provide an upper bound estimate that reveals the role of the substrate composition on qubit coherence.

In most heterostructures, $T_2$ is lower than that of the bare 2D host (Figure \ref{Fig:T2_Heterostructures}a) in agreement with a previous report.{\cite{Onizhuk2021_Substrate}} This indicates that the nuclear spin bath of the substrate material introduces additional noise that leads to faster qubit decoherence. However, we notice that $T_2$ is drastically lowered by certain substrates compared to others. For example, heterostructures of \ch{WS2} with \ch{CeO2} and CaO appear to maintain high $T_2$, almost equal to that of standalone \ch{WS2}, whereas $T_2$ is nearly an order of magnitude lower than the bare 2D host when \ch{WS2} is interfaced with other substrates, such as \ch{SiO2} (Figure \ref{Fig:T2_Heterostructures}b). We confirm that this is not due to the strain imposed on the 2D host material, since a strain of $\pm$5\% varies $T_2$ by no more than 20\% (\mytfigdone{Supplemantary Note 3}). Instead, not surprisingly, this can be explained by the $T_2$ values of the substrates themselves (Figure \ref{Fig:T2_Heterostructures}f). In particular, the $T_2$ of \ch{SiO2} (3.33 ms) is nearly an order of magnitude lower than that of \ch{CeO2} (45.7 ms) and CaO (31.9 ms), explaining the lower $T_2$ of heterostructures when \ch{SiO2} is the substrate material. (Figure \ref{Fig:T2_Heterostructures}b).

It is important to note that spin coherence is not always limited by the nuclear spins in the substrate material, and is instead limited by whichever component (substrate or 2D host material) is more noisy. This is clear in heterostructures of {\ch{Au2Se3O10}}, {\ch{Au2Se2O7}}, and {\ch{PdSO4}} with {\ch{CeO2}} and CaO (Figures {\ref{Fig:T2_Heterostructures}}c-{\ref{Fig:T2_Heterostructures}}e). Since $T_2$ of the substrates (Figure {\ref{Fig:T2_Heterostructures}}f) are nearly double that of the 2D host materials, the primary source of decoherence, not surprisingly, is the 2D material itself. As a result, from a materials design perspective, it is critical to select a 2D host material and substrate that both have high $T_2$ in order to maintain high $T_2$ in heterostructures.

We find that when qubit decoherence is driven by hyperfine interactions with nuclear spins, different facets of the substrate material yield similar $T_2$. For example, in \ch{WS2}/\ch{SiO2} heterostructures, $T_2$ for the (100), (010), and (001) planes of \ch{SiO2} are 4.06 ms, 4.07 ms, and 4.76 ms, respectively (Figure \ref{Fig:T2_Heterostructures}b). In fact, our results show that the nuclear spin bath-driven $T_2$ is relatively insensitive to the specific details of the heterostructural interface, whether it is confounded by the specific cleavage plane of the substrate, surface reconstruction, or even the interlayer distance between the 2D host and substrate material (more details are provided in \mytfigdone{Supplementary Note 4}). 
These results are likely valid only in the absence of other sources of noise that may be prevalent at the heterostructural interface, such as electric noise from dangling bonds.
%This is likely an artifact of neglecting other sources of noise that may be prevalent at the heterostructural interface, such as dangling bonds. 
Hence, because additional noise sources will lead to further decoherence, heterostructures with low predicted $T_2$ in our study are likely unsuitable candidates for hosting qubits with robust spin coherence.

\subsection{Models of coherence time $T_2$} \label{Sec:T2_Models}

Models of qubit coherence can greatly accelerate $T_2$ predictions. A recent work by Kanai et al.\cite{Kanai2022_ScalingLaw} proposed a model of $T_2$ based on scaling laws of isotopic features, namely the $g$-factor, nuclear spin number, and the isotope concentration. The scaling laws, which were determined for 3D materials, indicated that the contribution from a single nuclear isotope type, $i$, to the coherence time can be modeled as\cite{Kanai2022_ScalingLaw}
\begin{equation} \label{Eq:Kanai_T2_Isotope}
    T_{2,\mathrm{3D},i} = 1.5 \times 10^{18} |g_i|^{-1.6} I_i^{-1.1} n_i^{-1.0} (s)
\end{equation}
where $g_i$ is the $g$-factor, $I_i$ is the nuclear spin, and $n_i$ is the nuclear spin density. The overall $T_2$ of the material, assuming heteronuclear spin baths can be decoupled, is then calculated as
\begin{equation} \label{Eq:Kanai_T2}
    T_{2,\mathrm{3D}}^{-\eta} = \sum_i T_{2,\mathrm{3D},i}^{-\eta}
\end{equation}
where $\eta$ is the stretching exponent. Importantly, the model only requires the structure of the material and basic isotopic information, which can be found in e.g. the EasySpin database.{\cite{Stoll2006_EasySpin}} The model can therefore enable rapid predictions of $T_2$ for large sets of materials.{\cite{Kanai2022_ScalingLaw}}

\begin{figure}[!t]
    \centering
    \includegraphics[width=0.52\textwidth]{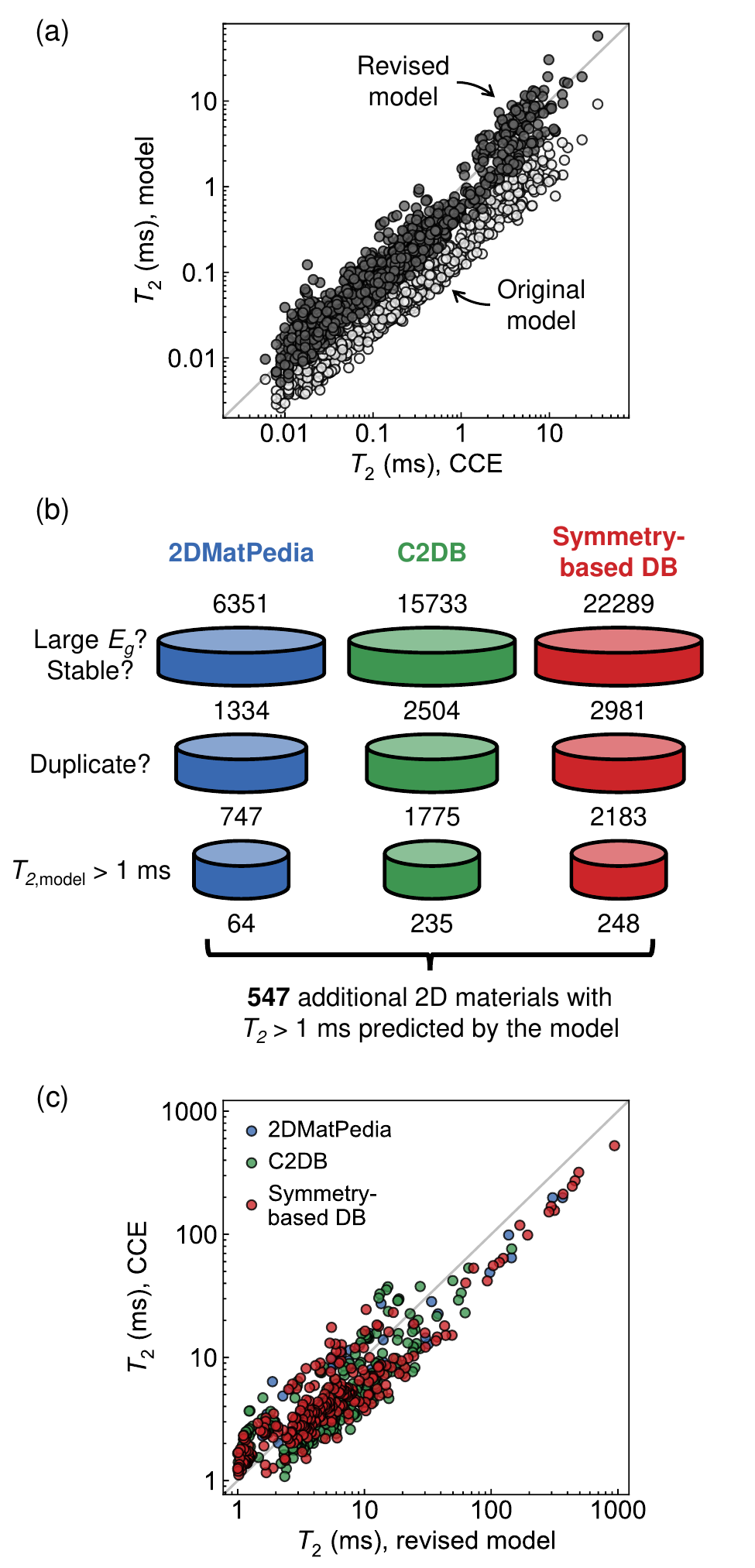}
    \caption{
    \textbf{Model of coherence time for 2D materials.}
    (a) Parity plot of the original (light) and revised (dark) models of the spin coherence time ($T_2$) for 2D materials.
    (b) Statistics from screening various 2D materials databases using the revised model of $T_2$.
    (c) $T_2$ predicted by the revised model for monolayers across various 2D materials databases. An additional \mytnote{546} monolayers with $T_1 > 1$ ms are found by screening these databases and verifying with CCE calculations.
    }
    \label{Fig:T2_Model}
\end{figure}

It is natural to ask whether the model, which was developed for 3D materials,\cite{Kanai2022_ScalingLaw} can also predict the $T_2$ times of 2D materials. We initially verify that heteronuclear spin baths can be decoupled in 2D systems (\mytfigdone{Supplementary Note 5}) similar to 3D systems,\cite{Seo2016_Decoherence, Ye2019_SpinCoherence_2D, Kanai2022_ScalingLaw} allowing us to test the model on 2D materials. However, using Eqs. (\ref{Eq:Kanai_T2_Isotope}) and (\ref{Eq:Kanai_T2}) and $\eta = 2$, following the methodology of Ref. \citenum{Kanai2022_ScalingLaw}, we find that the $T_2$ predictions by the model starkly deviate from those of CCE calculations (light dots in Figure \ref{Fig:T2_Model}a), indicating that a refined model is necessary.

\begin{figure}[!t]
    \centering
    \includegraphics[width=0.56\textwidth]{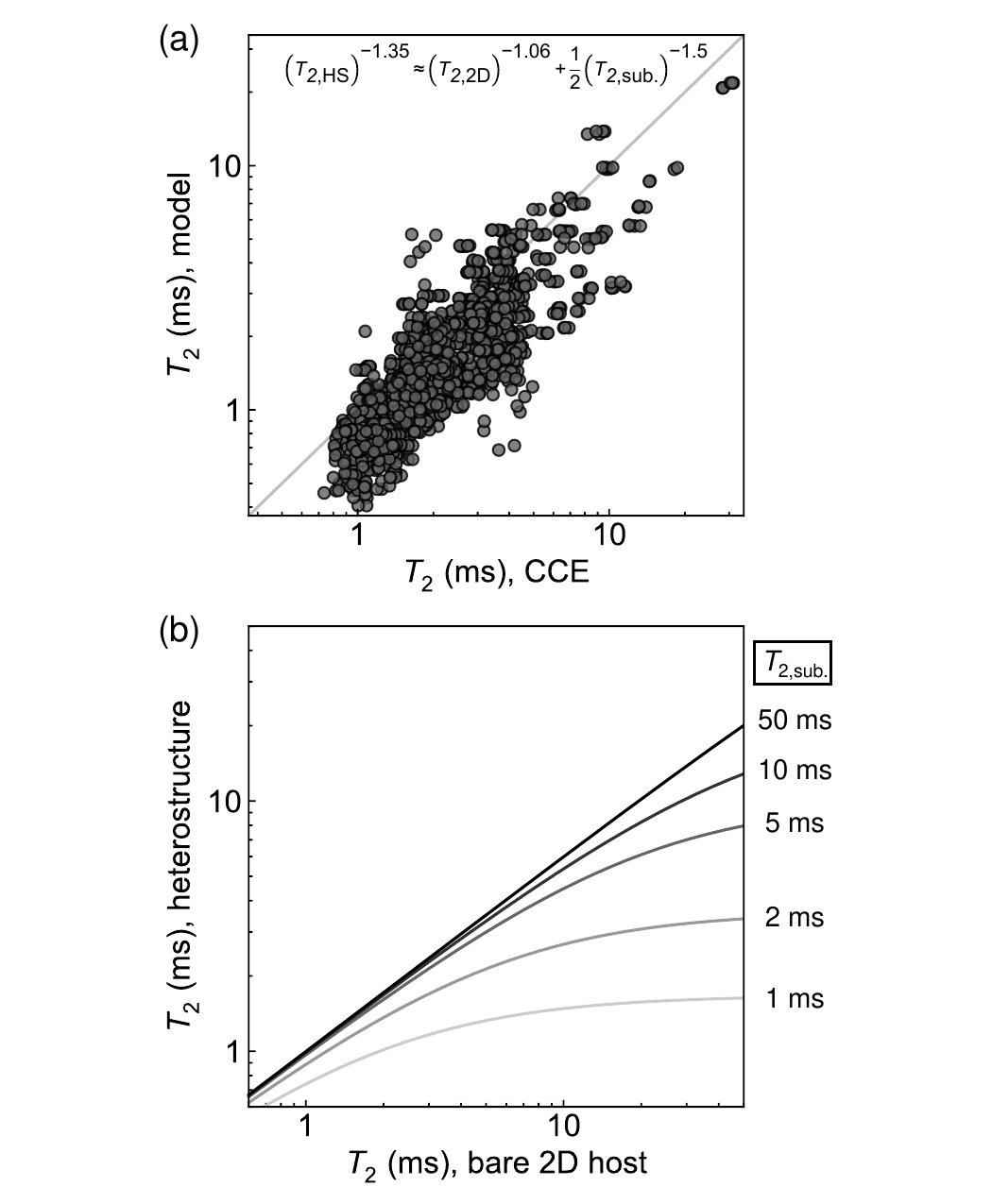}
    \caption{
    \textbf{Model of coherence time for heterostructures.}
    (a) The calculated coherence time ($T_2$) of heterostructures, compared to the $T_2$ predicted by the fitted model.
    (b) Model $T_2$ of heterostructures, calculated using Eq. (\ref{Eq:Model_T2_HS}), against $T_2$ of the bare 2D host material. Each curve corresponds to a different substrate $T_2$.
    }
    \label{Fig:T2_Model_Heterostructures}
\end{figure}

We find that a physically-motivated correction factor to the 3D model in Eqs. (\ref{Eq:Kanai_T2_Isotope}) and (\ref{Eq:Kanai_T2}) can enable better predictions of $T_2$ for 2D hosts. This is inspired by the semi-classical model of many-body spin dynamics proposed by Davis et al.\cite{Davis2023_CoherenceModel} where, importantly, information about the \emph{dimensionality} of the nuclear spin environment is encoded in the coherence function. As a result, we can derive a multiplicative factor that corrects for $T_2$ of a qubit in a 2D spin environment, given a model of $T_2$ parameterized for 3D spin environments (such as the model by Kanai et al.\cite{Kanai2022_ScalingLaw}). From the derivation provided in Supplementary Note 6, we arrive at the expression
\begin{equation} \label{Eq:Model_T2_2D_Isotope}
    T_{2,\mathrm{2D},i} = C(\alpha_\mathrm{2D})
    n_i^{(2-\alpha_\mathrm{2D})/3} w^{-\alpha_\mathrm{2D}/3}
    T_{2,\mathrm{3D},i} ,
\end{equation}
where $T_{2,\mathrm{2D},i}$ is the contribution from isotope $i$ to the coherence time, and $T_{2,\mathrm{3D},i}$ is given by Eq. (\ref{Eq:Kanai_T2_Isotope}). The correction factor is composed of a constant $C(\alpha_\mathrm{2D})$, the isotope concentration $n_i$ (in cm$^{-3}$), and the thickness $w$ of the 2D material (in cm). $\alpha_\mathrm{2D}$ represents the effective $1/r^\alpha$ interaction between the qubit and nuclear spins in 2D.\cite{Davis2023_CoherenceModel} Although dipolar interactions in 3D are described by $\alpha_\mathrm{3D} = 3$, we treat $\alpha_\mathrm{2D}$ as a fitting parameter due to the reduced screening in 2D. The $T_2$ for a qubit in a 2D host ($T_{2,\mathrm{2D}}$) can then be predicted using the formula
\begin{equation} \label{Eq:Model_T2_2D}
    T_{2,\mathrm{2D}}^{-3/\alpha_\mathrm{2D}} = \sum_i T_{2,\mathrm{2D},i}^{-3/\alpha_\mathrm{2D}} .
\end{equation}

Using the data generated from our high-throughput framework, we determine that $\alpha_\mathrm{2D} = 2.84$ yields the best fit to the CCE-calculated $T_2$ (dark dots in Figure \ref{Fig:T2_Model}a), resulting in the model
\begin{equation} \label{Eq:Model_T2_2D_Complete}
\begin{split}
    T_{2,\mathrm{2D},i} &\approx 0.94 n_i^{-0.28} w^{-0.95}
    T_{2,\mathrm{3D},i} \\
    T_{2,\mathrm{2D}}^{-1.06} &\approx \sum_i T_{2,\mathrm{2D},i}^{-1.06}
\end{split}
\end{equation}
The fact that the fitted $\alpha_\mathrm{2D}$ is slightly lower than the expected value from dipolar interactions in 3D ($\alpha_\mathrm{3D} = 3$) is consistent with the reduced screening in a 2D nuclear spin environment.\cite{Kezerashvili2022_Dipolar2D}

Using the model of $T_2$ in Eq. (\ref{Eq:Model_T2_2D_Complete}), we expand the search space for 2D qubit host materials by screening various databases, in particular 2DMatPedia,\cite{Zhou2019_2DMatPedia} C2DB,\cite{Haastrup2018_C2DB, Gjerding2021_C2DB} and a database constructed from symmetry considerations\cite{Wang2023_SymmBased} (Figure \ref{Fig:T2_Model}b). We initially screen materials that are stable, have wide band gap (>0.5 eV at the level of GGA), and are non-duplicate structures (\mytfigdone{Supplementary Note 6}). In total, we perform $T_2$ predictions on \mytnote{4705} non-duplicate 2D monolayers from the three databases, from which we identify \mytnote{547} host materials with $T_2 > 1$ ms (Figure \ref{Fig:T2_Model}b). We verify with CCE at order 2 that \mytnote{546} of these monolayers indeed exhibit $T_2 > 1$ ms (Figure \ref{Fig:T2_Model}c), demonstrating the ability to rapidly estimate $T_2$ on an expansive set of 2D host materials using the model in Eq. (\ref{Eq:Model_T2_2D_Complete}). We even identify some 2D hosts with $T_2 > 100$ ms through our analysis (Figure \ref{Fig:T2_Model}c), such as monolayer FeO and CaO. Note that our simple model only requires the structure of the 2D material and basic information about the nuclear isotopes.

We can also formulate a model of $T_2$ for heterostructures (denoted $T_{2,\mathrm{HS}}$) using the same semi-classical model of the coherence function proposed by Davis et al.{\cite{Davis2023_CoherenceModel}} (a full derivation is given in Supplementary Note 7). We arrive at the expression
\begin{equation} \label{Eq:Model_T2_HS}
    \left( \frac{1 (s)}{T_{2,\mathrm{HS}} (s)} \right)^{\eta_\mathrm{HS}}
    \approx \left( \frac{1 (s)}{T_{2,\mathrm{2D}} (s)} \right)^{3/\alpha_\mathrm{2D}} + \frac{1}{2} \left( \frac{1 (s)}{T_{2,\mathrm{sub.}} (s)} \right)^{3/2}
\end{equation}
where $T_{2,\mathrm{2D}}$ is the coherence time of the bare 2D material as given in Eq. (\ref{Eq:Model_T2_2D}), and $T_{2,\mathrm{sub.}}$ is the coherence time of the 3D substrate, which can be calculated using Eq. (\ref{Eq:Kanai_T2}). The ``(s)'' in each term represents the time unit, ensuring that the units are matched (\mytfigdone{Supplementary Note 7}). $\eta_\mathrm{HS}$ is the stretch exponent for heterostructures and is treated as a fitting parameter. By fitting the analytical model to our data, we find that
\begin{equation} \label{Eq:Model_T2_HS_Fitted}
    \left( \frac{1 (s)}{T_{2,\mathrm{HS}} (s)} \right)^{1.35}
    \approx \left( \frac{1 (s)}{T_{2,\mathrm{2D}} (s)} \right)^{1.06} + \frac{1}{2} \left( \frac{1 (s)}{T_{2,\mathrm{sub.}} (s)} \right)^{1.5}
\end{equation}
yields an appropriate fit of $T_2$ for heterostructures between a 2D host material and 3D substrate (Figure \ref{Fig:T2_Model_Heterostructures}a).

The form of Eq. (\ref{Eq:Model_T2_HS}) reflects the idea that the decoherence ``rate'' ($\sim 1/T^{\eta}$) of a qubit can be expressed as a sum of all sources of decoherence, in this case the 2D host material and substrate. The model also shows that $T_2$ of a heterostructure can be limited by either the 2D host or substrate, depending on which component has the noisier nuclear spin environment (i.e. has lower $T_2$). When $T_2$ of the 2D host is low, $T_2$ of the heterostructure is limited by the spin environment of the 2D host (Figure \ref{Fig:T2_Model_Heterostructures}b). On the other hand, when $T_2$ of the 2D host is high, $T_2$ of the heterostructure is limited by the substrate and therefore saturates with increasing $T_{2, \mathrm{2D}}$ (Figure \ref{Fig:T2_Model_Heterostructures}b). This result matches the findings from CCE simulations discussed earlier (Section \ref{Sec:Heterostructure_T2}).

\section{Discussion} \label{Sec:Discussion}

In this work, we develop a high-throughput computational workflow to survey the spin echo coherence times ($T_2$) of qubits in 2D materials and heterostructures. By applying the workflow to a set of \mytnote{1173} 2D materials, we identify \mytnote{190} monolayers with $T_2$ > 1 ms, including \ch{WS2}, \ch{Au2Se3O10}, \ch{Au2Se2O7}, and \ch{PdSO4}. We then apply the workflow to \mytnote{1554} lattice-matched heterostructures between high-$T_2$ 2D materials and substrates, where we find that placing 2D materials on low-noise substrates, such as \ch{CeO2} and CaO, can help maintain high $T_2$. In general, $T_2$ of a heterostructure can be limited by the nuclear spin bath of either the 2D host material or the substrate, depending on which component imposes a noisier nuclear spin environment. We then develop physically-motivated models of $T_2$ for 2D host materials and heterostructural systems, enabling rapid predictions based only on the structures of the 2D host and substrate materials.

To design qubit platforms with on-demand functionalities, spin defects must be identified within the host candidates. While this is beyond the scope of the present study, we can envision the possible electronic structures that can be realized through a simple symmetry analysis of structural motifs in the high-$T_2$ 2D host candidates. For example, \ch{Au2Se3O10}, \ch{Au2Se2O7}, and \ch{PdSO4} all possess transition metal complexes that are locally coordinated in a square-planar configuration (Figures \ref{Fig:T2_HT}c-e). Doping on the transition metal site, perhaps with another transition metal species with partially-filled $d$-orbitals, can potentially realize a spin-1 defect center, similar to the NV$^-$ center in diamond. This is because $d$-orbital splitting in a square planar coordination leads to the formation of doublet $e_g$ states, where two unpaired spins can form a spin-1 complex (\mytfigdone{Figure S8}). The rich chemical variety of 2D hosts elucidated by our study, and more importantly the diversity of locally-coordinated structural motifs, can enable the design of spin qubit-based devices tailored towards particular applications.

We expect that the large amount of data from our study, and future data that can be generated using the high-throughput CCE framework or the $T_2$ model presented here, can enable the generative design of materials to host spin qubits with robust coherence properties. Materials design supported by artificial intelligence has become widespread and has demonstratively benefited from large-scale databases, high-throughput computation, and even simple models. Various techniques to generate hypothetical 2D materials have already been explored,\cite{Song2021_Generative_2D, Lyngby2022_Generative_2D, Dong2023_Generative_2D} so perhaps analogous techniques can be developed for spin qubit applications by biasing these models towards materials with high $T_2$. Since the $T_2$ time is one of many properties in the complex, multivariate design space of spin qubit platforms, we envision that our work can be combined with other workflows to realize a comprehensive design scheme of qubit-based devices with functionalities tailored towards specific applications.

\section{Methods} \label{Sec:Methods}

\subsection{Cluster correlation expansion}

The decoherence dynamics of a spin qubit due to magnetic noise from surrounding nuclear isotopes (i.e., the nuclear spin bath) can be described by the many-body Hamiltonian
\begin{equation} \label{Eq:Hamiltonian}
    \hat{H} = \gamma_e \mathbf{B}\mathbf{S} + \sum_n \mathbf{B} \gamma_n \mathbf{I}_n + \mathbf{S} \mathbf{A}_n \mathbf{I}_n + \sum_{m < n} \mathbf{I}_n \mathbf{J}_{nm} \mathbf{I}_m .
\end{equation}
The first term represents the Zeeman splitting effect of the qubit spin $\mathbf{S}$ due to a magnetic field $\mathbf{B}$. The summation is performed over all nuclear spins $n$ with spin tensor $\mathbf{I}_n$. The second term represents the Zeeman splitting of the nuclear spins, the third term represents hyperfine interactions $\mathbf{A}_n$ between the qubit spin $\mathbf{S}$ and nuclear spins $\mathbf{I}_n$, and the fourth term represents nuclear dipole-dipole interactions represented by the tensor $\mathbf{J}_{nm}$. Here, we do not include the effects of zero-field splitting and quadrupolar interactions, which will be considered in future studies.

The coherence function $\mathcal{L}(t)$ of a spin qubit can be computed from the Hamiltonian in Eq. (\ref{Eq:Hamiltonian}), where
\begin{equation}
    \mathcal{L}(t) = \left| \frac{\langle 0 | \rho(t) | 1 \rangle}{\langle 0 | \rho(0) | 1 \rangle} \right|.
\end{equation}
Here, $\rho(t)$ is the time-dependent density matrix of the qubit spin. By applying a Hahn-echo pulse sequence, $T_2$ of a spin qubit can be measured and extracted by fitting $\mathcal{L}(t)$ to a stretched exponential function
\begin{equation}
    \mathcal{L}(t) = \exp\left[-\left(\frac{t}{T_2}\right)^n\right]
\end{equation}
where $n$ is a stretching exponent.

The dynamics of a qubit determined by its interaction with a nuclear spin bath can be simulated using the cluster correlation expansion (CCE) approach, which considers interactions between the qubit and clusters of nuclear spins in increasing cluster size (singlets, pairs, triplets, etc.). The CCE method is widely used to model decoherence dynamics of qubits in various systems, such as the NV$^-$ center in diamond and molecular color centers.\cite{Onizhuk2025_Review} Within the CCE approach, $\rho(t)$ is approximated by considering the dynamics of nuclear spin clusters with varying sizes, which allows $\mathcal{L}(t)$ to be factored into contributions from individual clusters:
\begin{equation}
    \mathcal{L}(t) = \prod_i \widetilde{\mathcal{L}}_{i}(t) \prod_{ij} \widetilde{\mathcal{L}}_{ij}(t) \prod_{ijk} \widetilde{\mathcal{L}}_{ijk}(t)...
\end{equation}
where $\widetilde{\mathcal{L}}_{i}(t)$, $\widetilde{\mathcal{L}}_{ij}(t)$, $\widetilde{\mathcal{L}}_{ijk}(t)$, etc. are the contributions from single spins $i$, pairs of spins $ij$, triplets of spins $ijk$, etc., respectively. It is important to note that $\mathcal{L}(t)$, and therefore $T_2$, can be measured using a Hahn-echo pulse sequence in experiments, providing a direct method to compare theoretical predictions with experiments.

\subsection{Computational details}

We used the PyCCE software\cite{Onizhuk2021_PyCCE} to run CCE simulations on 2D materials and heterostructures. In each system, we considered a model spin-1 defect, similar to the NV$^-$ center in diamond. Hyperfine interactions are treated using the point-dipole approximation. A magnetic field of 5 T is applied perpendicular to the plane of the 2D material in all calculations. The $g$-factor is assumed to be isotropic in our calculations. Due to the statistical nature of spin dynamics, we averaged over 40 CCE simulations for each system.

There are three principal user-defined parameters within PyCCE. First, interactions between the spin qubit and nuclear bath spins are calculated only when the bath spins are within a certain cutoff radius (r\_bath) from the qubit. Second, clusters are defined between nuclear bath spins that are within a cutoff radius (r\_dipole) from each other. Lastly, the CCE order sets the maximum size of the clusters included in the simulation; for example, calculations of order 3 (CCE-3) considers clusters of single, pairs, and triplets of nuclear bath spins. To obtain an accurate prediction of the $T_2$ time, it is imperative that CCE calculations are converged with respect to these three parameters. However, increasing r\_bath, r\_dipole, and order will also lead to longer computing time, because more nuclear bath spins are included in the calculation. This is especially a challenge when considering a large number of materials, each with their own unique set of nuclear isotopes and, therefore, different convergence criteria.

In order to strike a balance between accuracy and computing time, we employ a two-step strategy in our high-throughput workflow. First, we use a fixed, modest set of values for r\_bath (50 \AA), r\_dipole (15 \AA), and order (CCE-2) for all monolayers and predict their $T_2$ times. These parameters were benchmarked by Sajid and Thygesen.\cite{Sajid2022_2D_Coherence} We then rank the 2D materials based on their predicted $T_2$ times and perform proper convergence tests on those that exhibit $T_2$ > 1 ms. 
%The strategy rests on the hypothesis that using relatively low convergence parameters, which should exclude contributions from some nuclear spins, will establish an upper bound prediction of the $T_2$ time. In this way, our strategy should effectively mitigate false positive predictions, i.e. materials with high predicted $T_2$ time but low experimental $T_2$.
For all heterostructures, we use a fixed set of values for r\_bath (120 \AA), r\_dipole (30 \AA), and order (CCE-2) to predict $T_2$.

\subsection{Band gap predictions}

The band gap of a 2D material is a key indicator of its ability to host deep, localized defect states. To compute these gaps from first principles, we employed density functional theory (DFT),\cite{hohenberg1964inhomogeneous, kohn1965self, teale2022dft} which has proven highly successful at predicting ground-state properties across a wide range of materials. However, modeling the electronic structures of 2D systems remains challenging, since most exchange–correlation functionals are developed and benchmarked for bulk solids\cite{tran2009accurate} and may not perform reliably for low-dimensional systems.

To overcome this limitation, we adopted the screened-exchange range-separated hybrid (SE-RSH) functional,\cite{Zhan2023_SERSH, Zhan2025_SERSH_MetalOxides} which is capable of achieving good accuracy for both pristine and defective 2D materials.\cite{Zhan2023_SERSH} Within the SE-RSH approach, the exchange energy is expressed as a linear combination of exact (Hartree–Fock) and local exchange energies,\cite{ghosh2018combining} with the mixing fraction determined by a spatially-dependent local dielectric function. This formulation enables the SE-RSH functional to naturally capture complex screening variations in heterogeneous systems, including 2D materials, and their influence on the electronic structure.

The SE-RSH functional has been implemented in a modified version of the $\texttt{Qbox}$ code,\cite{gygi2008architecture} which employs a plane-wave basis set and norm-conserving pseudopotentials. We performed calculations on the relaxed 2D structures curated by the MC2D database without further structural relaxation, using supercells and $\Gamma$-point sampling, with optimized norm-conserving Vanderbilt pseudopotentials.\cite{hamann2013optimized} The energy cutoff for the plane-wave basis set was set at 90 Ry. To minimize the interactions between neighboring supercells, a minimum of 15$\text{\AA}$ vacuum spacing was included in our supercells.

\subsection{Note on hydrated systems}

In our high-throughput search of 2D materials with long $T_2$ times, we encountered some monolayers in which the experimentally-known 3D parent phase is naturally hydrated (i.e. have \ch{H2O} in their composition), but the hydrogen atoms are omitted from the structural CIF model likely because they could not be resolved using e.g. X-ray diffraction. This is especially prevalent for 3D layered systems, where there are well-defined spaces between the layers for \ch{H2O} to occupy and bind the layers. Such ``dehydrated'' structural models are problematic however, because they likely do not exist naturally. Moreover, because hydrogen can strongly decohere spin qubits (i.e. lower $T_2$) due to the high concentration of spinful nuclei, the absence of hydrogen atoms from the structure may lead to an overestimation of the $T_2$ time.
A representative example of this is talc, which has the natural chemical formula \ch{Mg3Si4O10(OH)2}. For the ``dehydrated'' structure \ch{Mg3Si4O12} (mc2d-471), we predict $T_2$ = 3.65 ms for the monolayer; however, for the true hydrated structure (mc2d-418), we predict $T_2$ = 0.085 ms for the monolayer, which is more than an of magnitude lower.

This requires us to check the original experimental studies of the 3D parent phases. Due to the large number of compounds considered in this study, we check by hand any 2D materials with predicted $T_2$ > 1 ms and omit those whose 3D parent phase is naturally hydrated. We do not check whether the parent phases of 2D materials with $T_2 < 1$ ms are naturally hydrated; in such cases, we expect $T_2$ to be lower than the predicted value for the dehydrated structure anyway.

\section*{Data and code availability}
Data and scripts can be found in Zenodo at \url{https://doi.org/10.5281/zenodo.16997329}.

\section*{Acknowledgements}

We thank Hosung Seo, Jonah Nagura, and Lien T. Le for helpful suggestions, as well as Nicola Marzari, Binbin Liu, and Kristjan Eimre for helpful discussions on the MC2D database.
This work was supported by the Midwest Integrated Center for Computational Materials (MICCoM). MICCoM is part of the Computational Materials Sciences Program funded by the U.S. Department of Energy, Office of Science, Basic Energy Sciences, Materials Sciences, and Engineering Division through the Argonne National Laboratory, under contract No. DEAC02-06CH11357.
This work was completed in part with resources provided by the University of Chicago’s Research Computing Center.
This research used resources of the National Energy Research Scientific Computing Center (NERSC), a DOE Office of Science User Facility supported by the Office of Science of the U.S. Department of Energy under Contract No. univers using NERSC award ALCC-ERCAP0025950.
%This work was partly supported by the Japan Society for the Promotion of Science (JSPS) Kakenhi Grant No. 23KK0092, Cross-ministerial Strategic Innovation Promotion Program, operated by the National Institutes for Quantum Science and Technology (QST), Precursory Research for Embryonic Science and Technology (PRESTO), Grant No. JPMJPR21B2, operated by the Japan Science and Technology Agency (JST), and Cooperative Research Projects of the Research Institute of Electrical Communication (RIEC), Tohoku University.
This work was partly supported by JSPS Kakenhi (Grant No. 23KK0092), QST Cross-ministerial Strategic Innovation Promotion Program, JST-PRESTO (Grant No. JPMJPR21B2), and RIEC Cooperative Research Projects.

\section*{Author Contributions}
M.Y.T.: methodology, software,formal analysis, investigation, data curation, writing (original draft), writing (editing). 
J.Z.: formal analysis, investigation, data curation, writing (editing).
S.K.: writing (editing).
G.G.: conceptualization, writing (editing), supervision, project administration, funding acquisition.
All authors contributed to the manuscript and approved the final version.

\section*{Competing Interests}
The authors declare no competing interests.

%----REFERENCES-----
\bibliography{biblio}
\bibliographystyle{rsc}
%---------------

\end{document}